\documentclass[conference]{IEEEtran}
\IEEEoverridecommandlockouts
\usepackage{amsmath,amssymb,amsfonts}
\usepackage{bm}
\usepackage{graphicx}
\usepackage[table]{xcolor}
\usepackage{cite}
\usepackage{booktabs}
\usepackage{enumitem}

\newtheorem{theorem}{Theorem}
\newtheorem{proposition}{Proposition}
\newtheorem{corollary}{Corollary}
\newtheorem{definition}{Definition}

\newcommand{\Zc}[1]{\mathcal{Z}_{#1}}
\newcommand{\Vc}[1]{\mathcal{V}_{#1}}
\newcommand{\one}{\mathbf{1}}
\newcommand{\E}{\mathbb{E}}
\newcommand{\R}{\mathbb{R}}
\newcommand{\Ktil}{\widetilde{K}}
\newcommand{\etat}[1]{\eta^{\mathrm{task}}_{#1}}
\newcommand{\spann}{\operatorname{span}}
\newcommand{\Var}{\operatorname{Var}}
\newcommand{\Cov}{\operatorname{Cov}}

\begin{document}

\title{Trustworthiness as a Measurement Structure for Task Conformance in 6G Service Chains}

\author{%
\IEEEauthorblockN{Bin Han\IEEEauthorrefmark{1}, Xintong Ling\IEEEauthorrefmark{2}, Muxia Sun\IEEEauthorrefmark{3}, and Hans D. Schotten\IEEEauthorrefmark{1}\IEEEauthorrefmark{4}}
\IEEEauthorblockA{\IEEEauthorrefmark{1}RPTU Kaiserslautern-Landau, Germany \quad
\IEEEauthorrefmark{2}Southeast University \& Purple Mountain Laboratories, China\\
\IEEEauthorrefmark{3}Beijing Huairou Laboratory, Beijing, China \quad
\IEEEauthorrefmark{4}German Research Centre for Artificial Intelligence (DFKI), Germany}}

\maketitle

\begin{abstract}
A 6G service crosses a chain of stages from different vendors, and its contract binds class-conditioned guarantees at the terminal interface: in the packet relay chain studied here, a priority class is owed a timely-delivery rate and a loss ceiling. We define trustworthiness as conformance of the delivered per-class distribution of outcomes to the contract and develop the measurement structure this forces. Two results fix its form: a discrepancy that treats outcomes interchangeably cannot express contract preferences, and no single expectation-valued score that determines the contracted terminal expectations can be carried from stage to stage once two or more contracted indicators remain independent at some stage. The minimal record is one signed number per independent contracted indicator, which the downstream design data carry to the terminal exactly; classical scalar trust models are the single-indicator case. The terminal deviation splits exactly into per-stage contributions. Transition logs at a stage estimate its contribution without reconstructing its full deployed behavior, a stage at its design behavior is flagged at most at a declared rate whatever traffic it receives, and the estimates bracket each stage's share of the terminal deviation at explicit confidence. In the three-relay chain, a relay that either deprioritizes premium traffic or has degraded capacity violates the same clause with identical contracted evidence: the transition audit localizes the relay in 300 of 300 windows and brackets its share of the deviation, and a diagnostic separates the causes.
\end{abstract}


\section{Introduction}\label{sec:intro}
A 6G network slice sold to a premium tenant traverses a multi-vendor chain: an access element, transport relays, and an edge processing stage, each specified by a different supplier and composed by an operator controlling none of their internals. IMT-2030 names trustworthiness among the defining requirements of 6G~\cite{itu2023imt}, and the conceptual literature attributes it to commitment (willingness) and competence (ability)~\cite{ling2025trust}. The operator needs this attribute quantitatively for three recurring decisions: whether a delivered service conforms to what a tenant bought, what each interface must report, and which vendor's stage introduced a deviation. Conformance is judged at the terminal interface where the contract binds, the other two decisions at individual stages. Vendors certify stages at design time, and the operator holds per-stage logs at deployment.

Trust has been quantified for networks along two lineages. Sun \emph{et al.} founded entropy-based trust for ad hoc networks, with concatenation and multipath rules for binary behaviors~\cite{sun2006trust}, and Theodorakopoulos and Baras formulated propagation as an axiomatic path algebra~\cite{theodorakopoulos2006trust}; Ling \emph{et al.} define trustworthiness as the trustee's probability of fulfilling a goal in a context~\cite{ling2025trust}. The unwilling/unable distinction has been operationalized for specific behaviors, malicious drops versus channel loss~\cite{shu2015privacy} and relay manipulation~\cite{graves2012detect}. These models evaluate a goal probability, whereas a 6G service-level agreement (SLA) is written per traffic class and outcome, so the deliverable at the terminal interface is the class-conditioned distribution of outcomes. None of these decomposes a multi-indicator contract deviation into per-stage contributions, separates unwillingness from inability under a general task specification, or attaches a stated confidence to such a decomposition.

The ingredients of a measurement that would close this gap exist separately. Reducing a stochastic model to the smallest linear state that reproduces prescribed outputs is classical, through reward-Krylov compression~\cite{poupart2002value}, exact constrained lumping~\cite{li1989lumping}, Koopman invariant subspaces~\cite{brunton2016koopman}, value equivalence~\cite{grimm2020value}, and rank converses~\cite{singh2004psr}; these compress a model for its outputs, while a service chain needs a record at every interface, carried backward through stage-varying design kernels and centered on the design. Additive propagation of per-stage model error is known as a Markov-chain perturbation inequality~\cite{mitrophanov2005sensitivity} and, for one scalar value functional, as an exact telescoping identity~\cite{yu2020mopo}, and the mismatch between a stage's deployed and nominal transition kernels, weighted by what the task values, appears as a training loss in value-aware model learning~\cite{farahmand2017value}. The bounds limit the terminal error without decomposing it, and the exact identities are scalar. Model-based fault diagnosis forms residuals against a nominal model~\cite{chow1984analytical} and tests them sequentially~\cite{basseville1993detection}, choosing residual directions that expose faults; here the terminal contract fixes the directions, and the propagated residuals sum to the terminal deviation the contract grades. Generic contraction coefficients~\cite{anantharam2013maximal,polyanskiy2017sdpi} quantify a channel's overall mixing whether or not it affects what a task depends on. Statistical model checking and conformance testing~\cite{ashok2019pac,gerhold2018conformance} and network tomography~\cite{duffield2006tomography} verify or localize at a set confidence, but full-kernel certificates carry alphabet-dependent cost.

We therefore define trustworthiness as conformance of the delivered per-class outcome distribution to the contract, an auditing problem as soon as stages are composed into a chain, and ask what a metric can be under this definition. A three-relay packet chain with a contracted priority class and a best-effort class is the running example. Propagation is the base operation of the measurement, as it is for scalar trust algebras: the record kept at a relay's output must reproduce the terminal key performance indicator (KPI) expectations through the design data of the relays downstream, and the deviation found there must separate into an inherited and an injected part. The contract must enter the measurement, and once it does, no scalar expectation record that determines the terminal KPIs survives composition when two KPI directions remain independent, so the propagated record is a vector.
\begin{enumerate}[leftmargin=1.35em]
\item \emph{A conformance specification for distribution deliverables}: trustworthiness is conformance of the delivered KPIs to the contract, and each ingredient of the measurement is either forced by a counterexample or declared a modeling choice.
\item \emph{A minimal propagating trust state}: one number per independent KPI direction at each hop reproduces every terminal expectation exactly, no smaller expectation record does, no scalar expectation record propagates once two KPI directions remain independent, magnitudes alone fix no floor verdict, and classical scalar models~\cite{sun2006trust,ling2025trust} are the single-KPI case.
\item \emph{Decisions from the state}: terminal deviation decomposes exactly into per-hop injections, the decomposition is not identifiable end to end over the unrestricted model class, and transition-level audits certify it at explicit confidence without estimating the deployed kernel, with a false-alarm bound that holds for every upstream input process. In the relay chain, policy non-compliance (a relay deprioritizing premium traffic) and capacity degradation, the unwilling and unable failures of trust models~\cite{ling2025trust,shu2015privacy}, produce identical contracted evidence; the audit localizes the relay and brackets its share, and a model-specific diagnostic separates the causes. Unlike a goodness-of-fit test, the audit ignores a relay change invisible to the contract.
\end{enumerate}

Although each mathematical component adapts an established construction, we are not aware of prior work that combines minimal expectation states induced by the terminal task, exact per-stage attribution of actual-versus-nominal mismatch, and confidence certificates of task dimension from transition logs. That combination is what the operator's three decisions require, and impossibility results force its base layers. The audited adversary is a vendor whose stage departs from its certified kernel through policy non-compliance, capacity degradation, or misconfiguration, with a fixed stochastic behavior over the audit window. Logs of both traffic classes, class labels, and episode identifiers are authenticated, so rewriting the class header is itself a logged violation. The audit protects a vendor at its certified kernel from blame for corrupted upstream traffic.

The remainder of the paper is organized as follows. Sec.~\ref{sec:spec} defines task-projected trustworthiness and what the measurement must read; Sec.~\ref{sec:state} constructs the minimal trust state and its calculus; Sec.~\ref{sec:decisions} derives the attribution identity and the transition-level certificate; Sec.~\ref{sec:casestudy} reports the case study; Sec.~\ref{sec:conclusion} concludes.

\begin{table}[t]
\renewcommand{\arraystretch}{1.15}
\setlength{\tabcolsep}{3pt}
\caption{Notation used throughout}
\label{tab:notation}
\centering
\footnotesize
\begin{tabular}{@{}ll@{\quad}ll@{}}
\toprule
$X\sim\mu$ & context (traffic class) & $Z_\ell$ & interface variable\\
$K_\ell,\Ktil_\ell$ & certified, deployed hop & $Q_\ell,P_\ell$ & nominal, actual marginal\\
$u_j,\mathcal C_x$ & utilities, admissible set & $\Phi_\ell,r_\ell$ & task basis, dimension\\
$\Delta_\ell$ & deviation state & $A_\ell$ & nominal transition\\
$\varepsilon_\ell,h_\ell$ & injected deviation & $\etat{\ell}$ & contraction per hop\\
$R^0_\ell$ & null threshold & $R^{\mathrm E}_\ell$ & estimation radius\\
\bottomrule
\end{tabular}
\end{table}

\section{The Specification Layer}\label{sec:spec}
A trustworthiness measurement reads the delivered conditional distribution, the nominal design it is judged against, and the contract. In the relay chain the trustee is a relay, or the chain up to an interface; its input is the traffic class together with the incoming packet state, its output the packet state it hands on, and the task is the SLA. In general, a \emph{trustee} is a stochastic system whose kernel $P_x$ is evaluated against a design-nominal kernel $Q_x$ through $m$ task utilities $\mathbf u_x$, delivering the KPIs $m_P(x)=\E_{P_x}[\mathbf u_x(Y)]$ at context $x$. Table~\ref{tab:notation} collects the symbols. The contract is a nonempty, closed, convex \emph{admissible set} $\mathcal C_x\subseteq\R^m$, whose constraints are floors $\E[u_j]\ge\theta_j(x)$ or ceilings, such as the timely-delivery and loss clauses, and tolerances $|\E[u_k]-\theta_k(x)|\le t_k(x)$ for deliverables where deviation in either direction is a defect.

\begin{definition}[Task-projected trustworthiness]\label{def:trust}
A trustee $(P,\mathcal C,\mu)$ is \emph{trustworthy for the task} if its delivered KPI vector is admissible, $m_P(x)\in\mathcal C_x$, at every context in the support of $\mu$. Its graded measurement is the \emph{signed KPI deviation} $\Delta_P(x)=m_P(x)-m_Q(x)$ from the design-nominal kernel and the \emph{compliance residual} $v_P(x)=m_P(x)-\Pi_{\mathcal C_x}m_P(x)$, the offset from the nearest admissible KPI vector, which vanishes exactly where the trustee is trustworthy.
\end{definition}

Trustworthiness is thus a property of the tuple $(P,\mathcal C,\mu)$, extending the task and context dependence of~\cite{ling2025trust} to the distributional setting. For per-clause constraints, $\mathcal C_x$ is a box, so $v_P$ lists the shortfall of each clause in its own units. 
\begin{proposition}[Output marginals do not suffice]\label{prop:necessity}
There exist service chains in which every output marginal coincides with its nominal counterpart while the conditional distribution violates its specification at every input.
\end{proposition}
\emph{Proof:} Let two classes of equal mass be served by the nominal chain from their own distributions and swapped by the deployed chain: every class-blind output marginal is unchanged, and each class receives the distribution of the other.

A per-input distance between the delivered and nominal distributions fails for a related reason. Under a $Q$-preserving relabeling $\tau$ of the outcome alphabet, every f-divergence, and every integral probability metric whose witness class is closed under composition with $\tau$, is invariant, $D(\tau_\#P\,\|\,Q)=D(P\,\|\,Q)$, while a task utility that is not $\tau$-invariant separates some $P$ from $\tau_\#P$. A witness class derived from the task utilities, as in integral probability metrics~\cite{muller1997ipm} and the discrepancy distance~\cite{mansour2009domain}, repairs this at a single stage; such a scalar is read from the state of Sec.~\ref{sec:state}, which by Theorem~\ref{thm:state}(ii) needs one expectation coordinate per independent KPI direction.

Scalars appear only at the boundary of the state: the compliance residual $v_P$ decides each clause, the contraction coefficient of Sec.~\ref{sec:state} measures downstream visibility, and the injection magnitude of Sec.~\ref{sec:decisions} localizes deviation to a stage.

Two ingredients are modeling choices: the admissible set, kept a set because contracts state floors, ceilings, and tolerances, and the class mix $\mu$, declared with the design and fixed in every offline quantity. Conformance is decided against $\mathcal C_x$ while the state is centered on $Q$; Theorem~\ref{thm:state} makes every terminal KPI affine in the state, and the design meets the contract, $m_Q(x)\in\mathcal C_x$, so zero deviation implies compliance.

\section{The Minimal Trust State and Its Calculus}\label{sec:state}
Relays are certified at design time, so the smallest record a relay's output must carry for the contract follows from design data alone. The chain has finite alphabets $\Zc{0},\dots,\Zc{L}$ and terminal utilities $\mathcal U_L=\{u_1,\dots,u_m\}$; distributions push forward along it, $Q_{\ell+1}=Q_\ell K_\ell$ nominally and $P_{\ell+1}=P_\ell\Ktil_\ell$ actually. Kernels act on functions by $(K_\ell f)(z)=\sum_{z'}K_\ell(z'|z)f(z')$, so $\E_{PK_\ell}[f]=\E_P[K_\ell f]$. Each interface variable carries the context beside the stage's payload, $Z_\ell=(X,S_\ell)$, and nominal and deployed kernels alike copy the context unchanged, as they copy the class header. Utilities $u_j(x,s)$ then encode per-input specifications, the cancellation of Proposition~\ref{prop:necessity} becomes impossible, and $\mu$ is the $X$-marginal of every stage distribution. With finitely many contexts of positive mass, the utilities $u_{x,j}(x',s)=\one\{x'=x\}\,u_j(x,s)/\mu(x)$ have unconditional expectations equal to the per-context KPIs, so Definition~\ref{def:trust} is decided coordinatewise from joint expectations, and $m$ counts this context-indexed family.

The distributions the state must serve form the \emph{admitted class} $\mathfrak P_\ell$: distributions on the \emph{certified domain}, the nominal support on which $K_\ell$ and the task functions are defined, with context marginal $\mu$; traffic outside the domain is itself nonconforming and shows directly in the logs. The interface variable is a sufficient stage state, memory such as a channel mode being folded into $\Zc{\ell}$.

\begin{definition}[Backward task spaces]\label{def:spaces}
Let $\mathcal G_X$ be the functions of the context alone, $\Vc{L}:=\spann(\mathcal G_X\cup\mathcal U_L)$ and, backward for $\ell=L-1,\dots,0$,
\begin{equation}\label{eq:backward}
\Vc{\ell}:=\spann\big(\mathcal G_X\cup K_\ell\,\Vc{\ell+1}\big),
\end{equation}
the stage-$\ell$ \emph{backward task functions}; unrolled, $\Vc{\ell}=\spann(\mathcal G_X\cup\{K_\ell\cdots K_{L-1}u_j\}_j)$, of dimension $|\mathcal X|+r_\ell$ with $r_0\le\cdots\le r_L\le m$ independent KPI directions beyond the context.
\end{definition}

Context functions carry no deviation over $\mathfrak P_\ell$, since every admitted distribution shares $\mu$, and $K_\ell\mathcal G_X\subseteq\mathcal G_X$ because the context is copied; with one context, $\mathcal G_X$ is the constants. We fix a \emph{conditionally centered} basis $\Phi_\ell$ of $\Vc{\ell}$ modulo $\mathcal G_X$ ($\E_{Q_\ell}[\Phi_\ell\,|\,X]=0$), let $\Sigma_\ell=\E_{Q_\ell}[\Phi_\ell\Phi_\ell^{\!\top}]$, the within-context covariance averaged over $\mu$, nonsingular on the certified domain, and define the \emph{deviation state} of an admitted distribution $P$ as $\Delta_\ell(P)=\E_P[\Phi_\ell]\in\R^{r_\ell}$, every task-relevant expectation recorded as an offset from the nominal design. In the relay chain $\Phi_\ell$ is the pair of class-$\mathrm p$ timely and loss indicators, each centered on its nominal class-$\mathrm p$ rate and divided by $\mu(\mathrm p)$, and $\Delta_\ell$ the pair of class-$\mathrm p$ KPI offsets at interface $\ell$, so $r_\ell=2$ at every relay output. Read at interface $\ell$, the relays upstream form one trustee in the sense of Definition~\ref{def:trust}, with task utilities $\Phi_\ell$, and $\Delta_\ell(P_\ell)$ is at once their trustworthiness record and the input deviation of the next relay.

\begin{theorem}[Exact propagation, minimal record]\label{thm:state}
(i) Unique matrices $A_\ell$ satisfy $K_\ell\Phi_{\ell+1}=A_\ell\Phi_\ell$, with no context-dependent offset. Consequently $\Delta_{\ell+1}=A_\ell\Delta_\ell$ under the nominal chain for every admitted input distribution, and every terminal utility expectation is an affine readout,
$\E_{PK_\ell\cdots K_{L-1}}[u_j]=\E_{Q_L}[u_j]+b_j^{\!\top}A_{L-1}\cdots A_\ell\,\Delta_\ell(P)$ with $b_j$ independent of $P$.
(ii) Any family of per-stage expectation records that determines the terminal expectations and updates stage to stage, for every admitted input distribution and any update rule, must, together with $\mathcal G_X$, span $\Vc{\ell}$; hence $r_\ell$ numbers are necessary as well as sufficient.
\end{theorem}
\emph{Proof:} (i) Each component of $K_\ell\Phi_{\ell+1}$ lies in $K_\ell\Vc{\ell+1}\subseteq\Vc{\ell}$, so $K_\ell\Phi_{\ell+1}=A_\ell\Phi_\ell+c_\ell(X)$ uniquely. Taking $\E_{Q_\ell}[\,\cdot\,|\,X=x]$, the left side is $\E_{Q_{\ell+1}}[\Phi_{\ell+1}\,|\,X=x]=0$ by the pushforward convention, applied per context, and conditional centering, and the right side is $c_\ell(x)$; hence $c_\ell=0$. The recursion follows from $\E_{PK_\ell}[\Phi_{\ell+1}]=\E_P[K_\ell\Phi_{\ell+1}]$, and iterating it gives the readout with $b_j$ the coordinates of $u_j$ in $\mathcal G_X\cup\Phi_L$, whose $\mathcal G_X$-part has the same expectation under every admitted distribution. (ii) If a record omits a direction $f$ of $\Vc{\ell}$ modulo $\mathcal G_X$, finite-dimensional separation gives a signed measure $\nu$ with zero mass in every context that annihilates the record but not $f$. Since $Q_\ell>0$ on the certified domain, $P^\pm=Q_\ell\pm t\nu$ are admitted for small $t>0$; they have equal records and different expectations of $f$, hence of some terminal utility, so no update rule exists.

In the relay chain the certified kernel of relay 3 thus determines both terminal KPIs exactly from the two offsets at its input, and Theorem~\ref{thm:state}(i) makes every compliance verdict computable from the state alone. A single number does not suffice: relay 3 carries the timely and loss offsets to the terminal separately, so one expectation at its input leaves a direction of the terminal KPIs undetermined.
\begin{corollary}[A single score carries at most one KPI]\label{cor:noscalar}
If some stage carries two or more independent KPI directions ($r_\ell\ge 2$), no scalar-valued expectation record that determines the terminal expectations for every admitted input distribution has a valid update rule. With a single terminal utility, $r_\ell\le1$ everywhere and the scalar deviation propagates exactly; scalar trust scores therefore compose exactly when the task they certify has a single KPI direction at every stage.
\end{corollary}

The quantitative cores of both trust lineages are this single-KPI case: the probability-based concatenation rule of Sun \emph{et al.}~\cite{sun2006trust}, $p_{12}=p_1p_2+(1-p_1)(1-p_2)$, is the scalar recursion with per-hop factor $2p-1$, since $2p_{12}-1=(2p_1-1)(2p_2-1)$; their entropy-based trust value is a bijection of $p$, and the measure of Ling \emph{et al.}~\cite{ling2025trust} is a one-utility specification. The sign that~\cite{sun2006trust} restores in dimension one remains necessary in every dimension: terminal timely-rate offsets of $\pm0.02$ have equal magnitude and only one violates the floor, and a hop that mixes coordinates maps equal magnitudes to unequal ones, so magnitudes in general admit no update rule.

The nominal design fixes how far a hop contracts task deviations. In the nominal covariance geometry the whitened state $d_\ell:=\Sigma_\ell^{-1/2}\Delta_\ell$, with symmetric roots, propagates by the whitened hop matrix $B_\ell:=\Sigma_{\ell+1}^{-1/2}A_\ell\Sigma_\ell^{1/2}$, and the \emph{task-restricted contraction coefficient}
\begin{equation}\label{eq:eta}
\etat{\ell}:=\|B_\ell\|_2^2
\end{equation}
is computable at design time from the certified kernels alone, with $\etat{\ell}\le1$ unconditionally: under the nominal chain $\E[\Phi_{\ell+1}(Z_{\ell+1})\,|\,Z_\ell]=(A_\ell\Phi_\ell)(Z_\ell)$, so the law of total variance gives $\Sigma_{\ell+1}=\E[\Cov(\Phi_{\ell+1}|Z_\ell)]+A_\ell\Sigma_\ell A_\ell^{\!\top}\succeq A_\ell\Sigma_\ell A_\ell^{\!\top}$. Equality $\etat{\ell}=1$ holds exactly when some task-visible observable, non-constant within contexts, passes the hop with zero conditional variance given the input. The generic squared maximal correlation and the Dobrushin coefficient bound it from above~\cite{anantharam2013maximal,polyanskiy2017sdpi}. In the relay chain both equal one, because the copied class is recoverable; with the context functions removed, the largest squared class-conditional maximal correlation and $\etat{}$ both equal $0.49$ at relay 2 and $0.88$ at relay 3, the class-$\mathrm p$ KPIs spanning every class-$\mathrm p$ state function.

\section{Decisions from the State}\label{sec:decisions}
The operator must decide which relay caused a terminal deviation and how confidently the logs support the answer.
\begin{theorem}[Mismatch-injection recursion]\label{thm:recursion}
For the deployed chain (arbitrary row-stochastic $\Ktil_\ell$; no closeness assumption),
\begin{equation}\label{eq:recursion}
\Delta_{\ell+1}=A_\ell\,\Delta_\ell+\varepsilon_\ell,\qquad
\varepsilon_\ell:=\E_{P_\ell}\!\big[(\Ktil_\ell-K_\ell)\Phi_{\ell+1}\big],
\end{equation}
and unrolling, with $A_{j:i}:=A_j\cdots A_i$ (the identity if $j<i$),
\begin{equation}\label{eq:attribution}
\Delta_L=A_{L-1:0}\,\Delta_0+\textstyle\sum_{\ell=0}^{L-1}A_{L-1:\ell+1}\,\varepsilon_\ell .
\end{equation}
\end{theorem}
\emph{Proof:} $\Delta_{\ell+1}=\E_{P_\ell}[\Ktil_\ell\Phi_{\ell+1}]
=\E_{P_\ell}[K_\ell\Phi_{\ell+1}]+\varepsilon_\ell=A_\ell\Delta_\ell+\varepsilon_\ell$.

The term $A_\ell\Delta_\ell$ is the deviation hop $\ell$ inherits, the corruption already present in its input, transported through its nominal behavior, and the \emph{injection} $\varepsilon_\ell$ its task-visible departure from its certified kernel, weighted by the traffic it receives. Each injection enters \eqref{eq:attribution} once, transported by the nominal hops downstream, so the interaction of two deviating hops falls to the downstream one. If a declared model mixes able and defecting kernels at stage $\ell$, $\Ktil_\ell=WK_{\mathrm{able}}+(1-W)K_{\mathrm{defect}}$ with willingness $W$, bilinearity splits $\varepsilon_\ell$ into inability despite goodwill and unwillingness-weighted defection, the unwilling/unable distinction of~\cite{ling2025trust}, which the case study instantiates as policy non-compliance and capacity degradation.

Downstream observation alone does not separate the two terms of \eqref{eq:recursion}: for any $\delta$ for which both pairs are feasible, the pairs $(\Delta_\ell,\varepsilon_\ell)$ and $(\Delta_\ell+\delta,\,\varepsilon_\ell-A_\ell\delta)$ produce identical downstream states. The ambiguity is realizable: on a binary alphabet a biased input distribution through a symmetric nominal hop and the nominal input distribution through a constant kernel of matching bias induce the same distribution from stage $\ell{+}1$ onward. Over the unrestricted chain model the inherited/injected decomposition is therefore not identifiable from end-to-end data at any sample size.

Transition-level data break the ambiguity: over one audit window the operator observes a preset number $n_\ell\ge2$ of transition pairs $(Z_{\ell,i},Z_{\ell+1,i})$ at each hop, one per service episode logged at every interface. The kernel $K_\ell$ and the task basis $\Phi_{\ell+1},\Sigma_{\ell+1}$ are known offline, and the whitened residuals $H_{\ell,i}=\Sigma_{\ell+1}^{-1/2}[\Phi_{\ell+1}(Z_{\ell+1,i})-(K_\ell\Phi_{\ell+1})(Z_{\ell,i})]$ average to $\hat h_\ell$, an unbiased estimate of the whitened injection $h_\ell:=\Sigma_{\ell+1}^{-1/2}\varepsilon_\ell$ computed without the deployed kernel. In whitened coordinates the recursion \eqref{eq:recursion} reads $d_{\ell+1}=B_\ell d_\ell+h_\ell$. The flag and the certificate use different radii. The \emph{null threshold} uses two constants per residual coordinate computed offline, the range $w_{\ell j}$ over all transition pairs of the certified domains and the variance proxy $v_{\ell j}=\max_z\Var(H_{\ell,j}\,|\,Z_\ell=z)$ under $K_\ell$, and with $L_\ell=\ln(2r_{\ell+1}/\alpha_\ell)$ for a false-alarm level $\alpha_\ell$
\begin{equation}\label{eq:radius}
R^0_\ell=\Big(\textstyle\sum_j t_{\ell j}^2\Big)^{1/2},\quad
t_{\ell j}=\tfrac{w_{\ell j}L_\ell}{n_\ell}+\Big(\big(\tfrac{w_{\ell j}L_\ell}{n_\ell}\big)^2+\tfrac{2v_{\ell j}L_\ell}{n_\ell}\Big)^{1/2},
\end{equation}
fixed before the logs; hop $\ell$ is flagged if $\|\hat h_\ell\|_2>R^0_\ell$. The \emph{estimation radius} $R^{\mathrm E}_\ell=(\sum_j(t^{\mathrm E}_{\ell j})^2)^{1/2}$ replaces the proxy by the unbiased sample variance $\hat v_{\ell j}$ of the logged residual coordinate, in the empirical-Bernstein form $t^{\mathrm E}_{\ell j}=(2\hat v_{\ell j}L'_\ell/n_\ell)^{1/2}+7w_{\ell j}L'_\ell/(3(n_\ell-1))$ with $L'_\ell=\ln(4r_{\ell+1}/\beta_\ell)$ for an estimation failure probability $\beta_\ell$~\cite{maurer2009empirical}; it depends on the logs and is valid under any deployed kernel for i.i.d. episodes. A rare contracted event or a class of small mass $\mu(x)$ inflates the whitened ranges, the latter roughly as $\mu(x)^{-1/2}$; both radii therefore carry variance terms.

\begin{theorem}[Certified localization and conformance]\label{thm:certificate}
(i) If hop $\ell$ draws each output from its certified kernel given its input and the window's history, then $\varepsilon_\ell=0$ identically, and the hop is falsely flagged with probability at most $\alpha_\ell$ for every upstream input process on the certified domain, including inputs chosen adaptively during the window, family-wise at most $\sum_\ell\alpha_\ell$. (ii) For i.i.d. episodes of one deployed chain, each logged at every interface, with probability at least $1-\beta_\ell$, $\|\hat h_\ell-h_\ell\|_2\le R^{\mathrm E}_\ell$, and on that event the flag fires whenever $\|h_\ell\|_2$ exceeds $R^0_\ell+R^{\mathrm E}_\ell$. (iii) For such episodes and a source at its design distribution, with probability at least $1-\sum_\ell\beta_\ell$ the whitened terminal deviation satisfies $\|d_L-\hat d_L\|_2\le\mathcal R:=\sum_\ell\big(\prod_{j>\ell}\sqrt{\etat{j}}\big)R^{\mathrm E}_\ell$, where $\hat d_L$ composes the estimates $\hat h_\ell$ through the whitened form of \eqref{eq:attribution}, and, with $c_\ell:=B_{L-1:\ell+1}^{\!\top}\Sigma_L^{1/2}b_j$, each hop's \emph{share} $c_\ell^{\!\top}h_\ell$ of a terminal utility deviation $b_j^{\!\top}\Delta_L$ lies within $a_\ell:=\sum_i|c_{\ell i}|\,t^{\mathrm E}_{\ell i}$ of $c_\ell^{\!\top}\hat h_\ell$ and the deviation within $\sum_\ell a_\ell\le\|\Sigma_L^{1/2}b_j\|_2\,\mathcal R$ of its estimate: a compliance certificate against any margin specification at explicit confidence.
\end{theorem}
\emph{Proof sketch:} (i) Under the certified kernel each residual coordinate is a bounded martingale difference with conditional variance at most $v_{\ell j}$ for any input sequence, so Freedman's inequality~\cite{freedman1975tail}, rescaled to the range $w_{\ell j}$, with a union bound over coordinates gives $\|\hat h_\ell\|_2\le R^0_\ell$ with probability at least $1-\alpha_\ell$. (ii) Under a deviating kernel the residual's conditional mean varies with the input, so the nominal proxy $v_{\ell j}$ need not bound its variance; the empirical-Bernstein inequality~\cite{maurer2009empirical}, applied two-sided per coordinate with the range $w_{\ell j}$, needs none, and the triangle inequality gives the flag. (iii) The whitened recursion is exact for any deployed kernels, and with $\|B_j\|_2=\sqrt{\etat{j}}$ the per-coordinate bounds of (ii) compose through the triangle inequality on their joint event.

Each hop reports $r_{\ell+1}\le m$ task coordinates, never the full $|\Zc{\ell}|\times|\Zc{\ell+1}|$ kernel. A goodness-of-fit test on the same logs also flags task-invisible deviation. With empirical input weights, the transition-count estimate of the full kernel gives the same $\hat h_\ell$, but certifying it needs intervals on every kernel entry and on the input frequencies, while the audit bounds only the residual average. At relay 2 with $n=10^5$ pairs, $R^0_\ell=0.027$ against an injection of norm $0.139$, while the range-only Hoeffding threshold is $0.242$ and detects nothing. With every episode logged at every interface the residual means telescope, $\hat d_L=\Sigma_L^{-1/2}\bar\Phi_L-B_{L-1:0}\Sigma_0^{-1/2}\bar\Phi_0$ for window means $\bar\Phi_\ell$, so the hop shares split this source-corrected sample estimate exactly. A direct terminal bracket is narrower but assigns no share to a hop: terminal samples decide each clause by the side of its floor or ceiling on which the whole bracket lies, and hop logs localize the deviation. A fired flag rejects at level $\alpha_\ell$ the hypothesis that the hop runs its certified kernel, which end-to-end data cannot test, and on the estimation event an unfired flag bounds the local injection by $R^0_\ell+R^{\mathrm E}_\ell$.

\begin{figure*}[t]
\centering
\includegraphics[width=\textwidth]{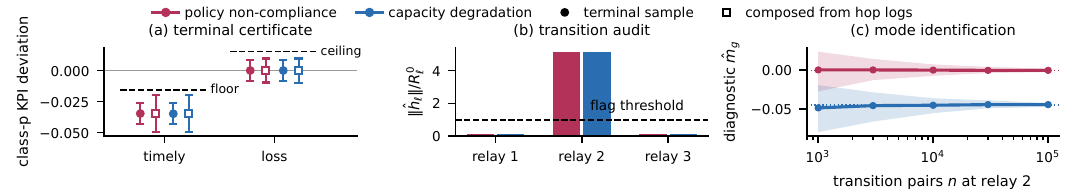}
\caption{(a) Terminal certificate for the class-$\mathrm p$ KPIs ($n=10^5$, $300$ windows): estimates from the terminal samples and composed from the hop logs of the same episodes, with family-wise $95\%$ half-widths, coordinatewise as in Theorem~\ref{thm:certificate}(iii); means over windows of the per-window values; dashed: SLA floor and ceiling. (b) Transition audit (family-wise $95\%$): relay 2 flagged in $300/300$ windows per mode, relays 1 and 3 in no window. (c) Mode identification: $\hat m_g$, $120$ windows per grid point (bands $5$--$95\%$; dotted: predictions).}
\label{fig:audit}
\end{figure*}

\section{Case Study: Multi-Class Packet Relay Chain}\label{sec:casestudy}
In the relay chain, packets of the priority class $\mathrm p$ and the best-effort class $\mathrm b$ arrive in the mix $\mu=(0.3,0.7)$ and cross three relays operated by different vendors. The class, a header field copied through every hop, is the context $X$; the packet state $S_\ell\in\{\mathrm D,\mathrm L,\mathrm E\}$ records whether the packet is still deadline-feasible, already late, or lost, with $\mathrm E$ absorbing and $\mathrm L$ never returning to $\mathrm D$. Each nominal relay loses a packet with probability $0.0040$, $0.0091$ and $0.0018$ (relays 1 to 3) and delays a surviving deadline-feasible packet past its deadline with probability $0.01$ for class $\mathrm p$ and $0.10$ for class $\mathrm b$; these numbers fix every kernel row. The premium tenant's SLA binds $\theta_{\mathrm D}=0.94$ and $\theta_{\mathrm E}=0.03$ for class $\mathrm p$; the nominal design delivers $0.956$ and $0.015$. No scalar trust score propagates the contracted KPIs: two admitted distributions at the output of relay 2 with the same class-$\mathrm p$ timely rate $0.932$ and loss rates $0.013$ and $0.023$ reach terminal loss rates $0.015$ and $0.025$, which no record of the timely rate distinguishes. At the ingress, where every function on the certified domain $\{(x,\mathrm D)\}$ is a context function, the constants-only construction would report a spurious coordinate with $\etat{}=0.99$, while the construction modulo $\mathcal G_X$ reports $r_0=0$.

Relay 2 deviates in two modes constructed to coincide in every contracted direction. Under policy non-compliance the vendor deprioritizes premium traffic, handling a class-$\mathrm p$ packet as best effort with probability $0.4$, $\Ktil=0.6\,K+0.4\,K_{\mathrm{defect}}$, so class-$\mathrm p$ lateness becomes $0.6(0.01)+0.4(0.10)=0.046$. Under capacity degradation lateness rises for both classes by $0.036$, to $0.046$ and $0.136$, so the class-$\mathrm p$ rows of the two deployed kernels coincide exactly. Both modes deliver the same contracted evidence: the class-$\mathrm p$ timely rate falls to $0.921$, below the floor by a compliance residual of $0.019$, while the loss clause stays met at $0.015$. A class-blind timeliness monitor registers a drop from $0.790$ to $0.779$ under non-compliance and to $0.759$ under degradation, the premium violation diluted by aggregation over classes. With relay $k$ as hop $\ell=k-1$, the state at the output of relay 2 is $\Delta_2=(-0.0352,\,0)$ in KPI units, and by the attribution identity \eqref{eq:attribution} the terminal deviation is this injection transported through the nominal relay 3, $\Delta_L=A_2\varepsilon_1=(-0.0348,\,0)$, the identity holding to machine precision. An injection at relay 1 with the same contracted terminal effect would produce the same contracted terminal state, so terminal KPI samples cannot identify the deviating relay.

Fig.~\ref{fig:audit} reports the audit with $n=10^5$ episodes per window, about $3\times10^4$ of them in class $\mathrm p$, and $300$ independent windows per mode, with $\alpha_\ell=\beta_\ell=0.05/3$. The transition audit localizes the injecting relay without specifying an alternative failure model (Fig.~\ref{fig:audit}b). With the null threshold \eqref{eq:radius}, relay 2 is flagged in $300$ of $300$ windows per mode ($95\%$ Wilson interval $[0.987,1]$) and relays 1 and 3 in none (per-window false-flag probability below $0.010$, one-sided $95\%$ Clopper--Pearson), although relay 3 carries the corrupted traffic, consistent with Theorem~\ref{thm:certificate}(i). A third deviation, raising relay 2's best-effort lateness from $0.10$ to $0.20$, cuts best-effort timely delivery by $0.080$ but its injection is exactly zero. A Pearson chi-square test of the same transition pairs against the certified kernel, at the audit's level and conditional on input counts, detects it in all $300$ windows, the transition audit in none, while both tests fire in every window of the two violating modes. The estimation radius at relay 2 stays at or below $R^{\mathrm E}_\ell=0.027$ in every window, so the power margin $\|h_\ell\|_2-R^0_\ell-R^{\mathrm E}_\ell$ exceeds $0.085$ in all $300$ windows of both modes. Since $r_0=0$, in every window the composed estimate equals the class-$\mathrm p$ terminal sample deviation scaled by $n_{\mathrm p}/(n\mu(\mathrm p))$ for the realized class-$\mathrm p$ count $n_{\mathrm p}$. Its bracket (Fig.~\ref{fig:audit}a), $\pm0.015$ on the timely rate with relay 2's radius scaled by $(\etat{2})^{1/2}=0.94$, covers the true value in $300$ of $300$ windows and lies below the floor in $300$ windows under non-compliance and $299$ under degradation; the direct bracket from the same terminal samples, $\pm0.009$, does so in all $300$. In every window relay 2's share bracket lies below zero, the brackets of relays 1 and 3 contain zero, and all three cover the true shares; the mean brackets are $-0.035\pm0.007$, $\pm0.003$ and $\pm0.004$.

With both kernels named, the mode is decided afterwards from relay 2's transition pairs: the window average $\hat m_g$ of the residual in the direction $g=\one\{X=\mathrm b\}(\one\{S=\mathrm D\}-\one\{S=\mathrm L\})$, where the two kernels differ, assigned to the nearer predicted value, identifies the mode in $300$ of $300$ windows at $n=10^5$ and in $120$ of $120$ from $n=10^4$ (Fig.~\ref{fig:audit}c). The contracted state is identical at every interface, so no KPI audit separates the modes at any sample size. On full logs of both classes the Chernoff information $C(P,Q)=-\min_{s\in[0,1]}\log\sum_zP(z)^sQ(z)^{1-s}$, the optimal error exponent for equal priors, is $9.7\times10^{-4}$ for transition pairs at relay 2 and $3.6\times10^{-4}$ for terminal states, so terminal diagnosis needs $2.7$ times the samples asymptotically. For identical contracted evidence the remedy differs: the operator renegotiates with or penalizes the transport vendor under non-compliance and replaces or upgrades the relay under degradation. All values come from fixed-seed scripts.

\section{Conclusion}\label{sec:conclusion}
Three impossibility results force the base of the measurement: relabeling-invariant discrepancies cannot express task orderings, scalar expectation records that determine two independent terminal KPIs do not survive composition, and magnitude-only records fix no floor verdict and in general do not propagate. The minimal record they leave is a signed state that propagates exactly. On this structure the operator obtains exact attribution of terminal deviation to hops, design-time contraction coefficients, and two guarantees against the certified kernels: a false-alarm bound for a vendor at its certified kernel that holds for every upstream input process, and, for i.i.d. episodes under any deployed kernels, brackets on the hop shares, which split the source-corrected terminal sample estimate exactly.

The chain calculus is the base case of a network trust calculus: multipath and recommendation fusion are an estimation layer on the same state, and the willingness mixture interfaces with the expectation-maximization estimation of Ling \emph{et al.}~\cite{ling2025trust}. Stages that adapt to the audit or confine deviation to rarely visited states, colluding logs, errors in the certified kernels, sequential audits, and larger state alphabets are left for future work.

\bibliographystyle{IEEEtran}
\bibliography{references}

\end{document}